\documentclass[12pt,a4]{article}
\usepackage{setspace}
\usepackage{amssymb}
\usepackage{amsmath,amssymb}
\usepackage[T1]{fontenc}
\usepackage[bf,small,tableposition=top]{caption}
\usepackage{subfig}
\usepackage{amsthm}
\usepackage{lineno}
\usepackage[figuresright]{rotating}
\usepackage{enumerate}
\usepackage{authblk}
\usepackage[T1]{fontenc}
\usepackage[utf8]{inputenc}
\usepackage{graphics}
\usepackage{graphicx}
\usepackage{multirow}
\usepackage{lscape}
\usepackage[dvipsnames]{xcolor}
\usepackage{mathrsfs}
\usepackage{hyperref}
\usepackage{xcolor}
\usepackage{textcomp}
\usepackage{listings}
\usepackage[dvipsnames]{xcolor}
\begin{document}
	\title{ Negative binomial-reciprocal inverse Gaussian distribution: Statistical properties with applications}
	\author{Ishfaq S. Ahmad$^{1}$ , Anwar Hassan$^{2}$ and Peer Bilal Ahmad$^{3\dag}$\\
	
	$^{1,2}$  P. G.Department of Statistics, University of Kashmir, Srinagar, India \\
	$^\dag$ Corresponding author- peerbilal@yahoo.co.in \\
	$^{3}$ Department of Mathematical Sciences, IUST, Srinagar, India
}

\maketitle

\begin{abstract}
	In this article, we propose a new three parameter distribution by compounding negative binomial with reciprocal inverse Gaussian model called negative binomial - reciprocal inverse Gaussian distribution.   This model is tractable with some important properties not only in actuarial science but in other fields as well where overdispersion pattern is seen. Some basic properties including recurrence relation of probabilities for computation of successive probabilities have been discussed. In its compound version, when the claims are absolutely continious, an integral equation for the probability density function is discussed. Brief discussion about extension of univariate version have also been done to its respective multivariate version. parameters involved in the model  have been estimated by Maximum Likelihood Estimation technique. Applications of the proposed distribution are carried out by taking two real count data sets. The result shown that the negative binomial- reciprocal inverse Gaussian distribution gives better  fit when compared  to the Poisson and negative binomial distributions.     
\end{abstract}


\section{Introduction}
 Many researchers  often encounters such practical situations that involve count variables. A count variable can only take on positive integer values or zero because an event cannot occur a negative number of times. There are numerous examples of count data, for example, the number of insurance claims, the number of accidents on a particular busy road crossing, number of days a patient remains admitted in a hospital, the number of alcoholic drinks consumed per day (Armeli et al., 2015), the number of cigarettes smoked by adolescents (Siddiqui, Mott, Anderson and Flay, 1999) and so on. Undoubtedly, the one-parameter Poisson distribution is the most popular model for count data used in practice, mainly because of its simplicity. A major drawback of this distribution is that its equidispersion property, i.e.,  the variance is equal to the mean. Count data often exhibit underdispersion or overdispersion. Overdispersion relative to the Poisson distribution is when the sample variance is substantially in excess of the sample mean. Underdispersion relative to the Poisson is when the sample mean is substantially  in excess of the sample variance.   \par 
 Many attempts have been made to develop such models that are less restrictive than Poisson, and are based on other distributions, have been presented in the statistical literature, including the negative binomial, generalized Poisson and generalized negative binomial (see Cameron and Trivedi (1998) and Famoye (1995), among others). Also various methods have been employed to develop new class of discrete distributions like  mixed Poisson method (see Karlis and Xekalaki, 2005), discretization of continuous family of distribution and discrete analogues of continuous distribution. \par 
 Mixture approach is one of the prominent method of obtaining new probability distributions in applied field of probability and statistics, mainly because of its simplicity and unambiguous interpretation of the unobserved heterogeneity that is likely to occur in most of practical situations. In this paper a negative binomial (NB) mixture model that includes as mixing distribution the reciprocal inverse Gaussian distribution is proposed by taking  $p=exp(-\lambda)$,( where $p$ is negative binomial parameter) assuming that $\lambda$ is distributed according to a
 reciprocal inverse Gaussian distribution, obtaining the negative binomial-reciprocal inverse Gaussian distribution, denoted by  $\mathcal{NBRIG}$, which can be viewed as a comparative model to negative binomial distribution and Poisson distribution. \par  
 The new distribution is unimodal, having thick tails, positively or negatively skewed and posses over-dispersion character. Recursive expressions of probabilities are also obtained which are an important component in compound distributions particularly in collective risk model. Basically there are three parameters involved in the new distribution which have been estimated by using an important technique namely Maximum Likelihood Estimation(MLE) and goodness of fit has been checked by using chi-square criterion. \par 
 The rest of the paper is structured as follows: In Section 2, we study some basic characteristics of the distribution like probability mass function (PMF), PMF plot, factorial moments and over-dispersion property. In section 3, we study  $\mathcal{NBRIG}$ as compound distribution and recurrence relation of probabilities are being discussed to compute successive probabilities. Extension of univariate to multivariate version  have been discussed briefly in section 4. Section 5 contains information about estimation of parameters by MLE. Two numerical illustrations have been discussed in section 6 followed by conclusion part in section 7.

\section{Basic Results}
In this section we introduce the definition and some basic statistical properties of $\mathcal{NBRIG}$ distribution. But we will start with classical negative binomial distribution denoted as $ X\sim \mathcal{NB}(r,p)$ whose probability mass function given by:
\begin{equation}
P(X=x)=\binom{r+x-1}{x}p^r q^x , \quad x=0,1,\cdots 
\end{equation}
denoted as $ X\sim \mathcal{NB}(r,p)$ with $r>0$, $q=1-p $ and $0<p<1$. Since its usage is important later, so we will discuss some important characteristics of this distribution. The first three moments about zero of $\mathcal{NB}(r,p)$ distribution are given by:
\begin{equation}
\begin{aligned} 
\mathbb{E}(X)=& \frac{r(1-p)}{p} \\ \nonumber
\mathbb{E}(X^2)=& \frac{r(1-p)\left[1+r(1-p)\right]}{p^2} \\ \nonumber
\mathbb{E}(X^3)=& \frac{r(1-p)}{p^3}\left[1+(3r+1)(1-p)+r^{2}(1-p)^2\right] \nonumber
\end{aligned}
\end{equation}
Also the factorial moment of $\mathcal{NB}(r,p)$ distribution of order $k$ is:
\begin{equation}
\begin{aligned} 
\mu_{[k]}(X)=& \mathbb{E}\left[X(X-1)\cdots(X-k+1)\right] \\ 
=& \frac{\Gamma(r+k)}{\Gamma(r)}\frac{(1-p)^k}{p^k}, \quad k=1,2,\cdots
\end{aligned}
\end{equation}
Let random variable $Z$ has reciprocal inverse Gaussian distribution whose probability density function is given by 
\begin{equation}
f(z,\alpha,m)=\sqrt{\frac{\alpha}{2\pi z}}e^{-\frac{\alpha}{2m}\left(zm-2+\frac{1}{zm}\right)}  ,\quad z>0
\end{equation}
where $\alpha$, $m>0$. We will denote $ Z\sim \mathcal{RIG}(\alpha,m)$. The moment generating function (mgf) of $\mathcal{RIG}(\alpha,m)$ is given by:
\begin{equation} \label{mxt}
M_{Z}(t)=\sqrt{\frac{\alpha}{\alpha-2t }} exp\left\{\frac{\alpha}{m^2}\left[m-\frac{m}{\sqrt{\alpha}}\sqrt{\alpha-2t}\right]\right\}.
\end{equation}
\noindent \textbf{Definition 1.} A random variable $X$ is said to have negative binomial -reciprocal inverse Gaussian distribution if it follows the stochastic representation as:
\begin{equation}  \label{sr}
\begin{aligned}
X|\lambda\sim  &\mathcal{NB}(r,p=e^{-\lambda})\\
\lambda \sim  &\mathcal{RIG} (\alpha,m)
\end{aligned} 
\end{equation}
\textit{where} $r, \alpha,m>0$ \textit{and we can write  $X  \sim \mathcal{NBRIG}(r,\alpha,m)$} and is obtained in Theorem 1. \\
\noindent \textbf{Theorem 1.} \textit{Let $X  \sim \mathcal{NBRIG}(r,\alpha,m)$ be a negative binomial -reciprocal inverse Gaussian distribution as defined in (\ref{sr}) then PMF is given by}
\begin{equation} \label{pmf} 
p(x)=\binom{r+ x-1}{x}\sum_{j=0}^{x}\binom{x}{j}(-1)^{j}\sqrt{\frac{\alpha}{\alpha+2(r+j) }} exp\left\{\frac{\alpha}{m^2}\left[m-\frac{m}{\sqrt{\alpha}}\sqrt{\alpha+2(r+j)}\right]\right\}, 
\end{equation}
\textit{with $ x=0,1,\cdots$ and  $ r, \alpha, m>0$.}\\

\noindent \textit{Proof:} Since $ X|\lambda \sim \mathcal{NB}(r,p=e^{-\lambda}) $ and $\lambda \sim  \mathcal{RIG} (\alpha,m)$. Then unconditional PMF of $X$ is given by 
\begin{eqnarray} \label{uc}
p\left(	X=x\right)&=&\int_{0}^{\infty}f(X|\lambda)g(\lambda;\alpha,m)d\lambda   
\end{eqnarray}
where 
\begin{eqnarray} \label{nb}
f(x|\lambda)&=&\binom{r+ x-1}{x} e^{-\lambda r}(1-e^{-\lambda})^x \nonumber \\
&=&\binom{r+ x-1}{x}\sum_{j=0}^{x}\binom{x}{j}(-1)^{j}(-1)^j e^{-(r+j)}
\end{eqnarray}
and $g(\lambda;\alpha,m)$ is the probability density function(pdf) of  $\mathcal{RIG} (\alpha,m)$.\\
Put (\ref{nb}) in  Equation (\ref{uc}), we get
\begin{eqnarray} \label{mgf}
p\left(	X=x\right)&=&\binom{r+ x-1}{x} \sum_{j=0}^{x}\binom{x}{j}(-1)^{j}\int_{0}^{\infty}e^{-(r+j)}g(\lambda;\alpha,m)d\lambda \nonumber \\
&=&\binom{r+ x-1}{x} \sum_{j=0}^{x}\binom{x}{j}(-1)^{j} M_{\lambda}\left(-(r+j)\right) 
\end{eqnarray}
Use (\ref{mxt}) in Equation (\ref{mgf}) to get 	PMF of  $\mathbb{NBRIG}(r,\alpha,m)$ as
\begin{equation*}
p(x)=\binom{r+ x-1}{x}\sum_{j=0}^{x}\binom{x}{j}(-1)^{j}\sqrt{\frac{\alpha}{\alpha+2(r+j) }} exp\left\{\frac{\alpha}{m^2}\left[m-\frac{m}{\sqrt{\alpha}}\sqrt{\alpha+2(r+j)}\right]\right\}, 
\end{equation*} 
which proves the theorem. \hfill $\blacksquare$ \\
\begin{figure}[htp]
	\begin{center}
		\includegraphics[scale=0.5]{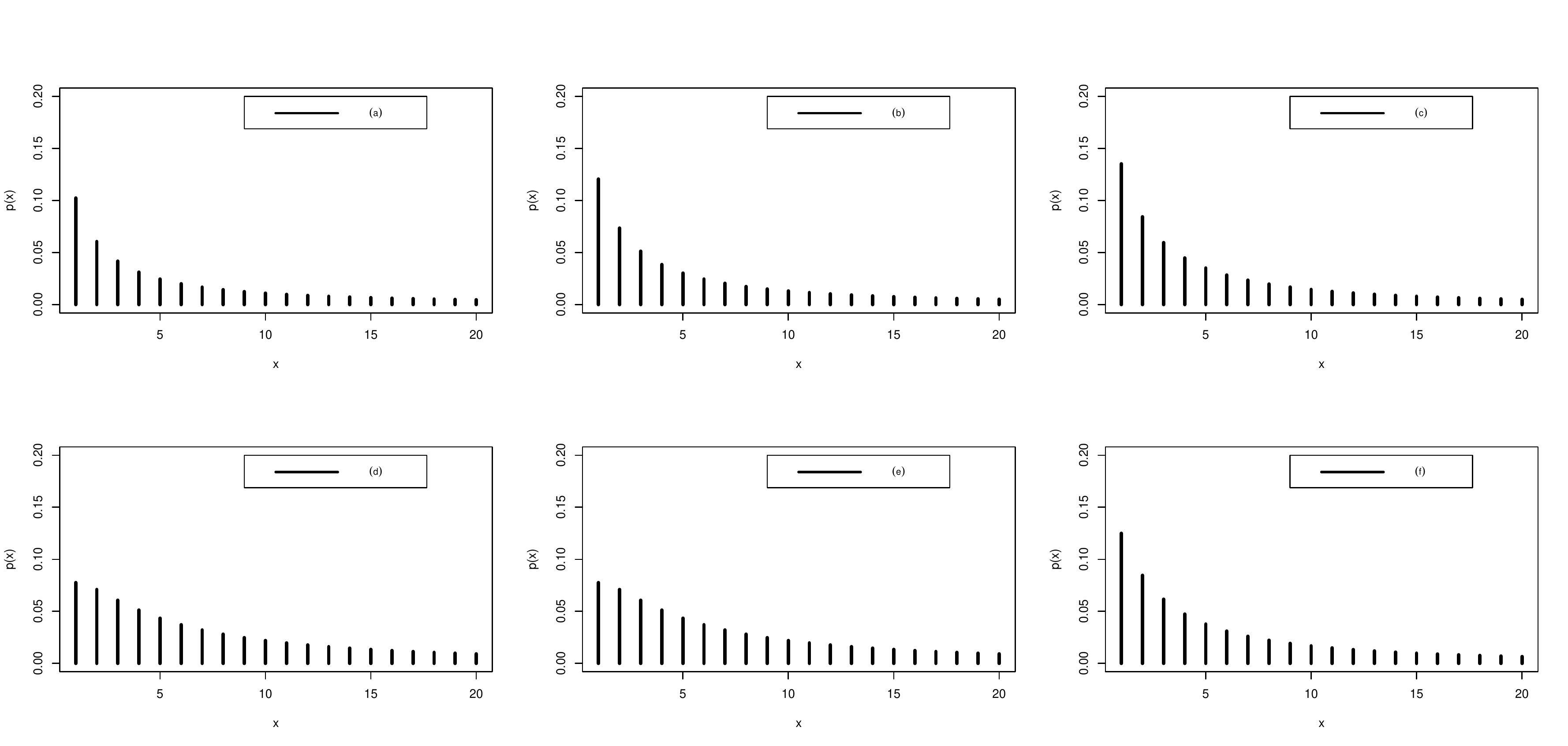}
		\captionof{figure}{PMF plot of $\mathcal{NBRIG}(r,\alpha,m)$ distribution for different valaues of parameters: (a) $r=0.5,\alpha=0.5,m=0.5$, (b)$r=0.5,\alpha=1,m=0.5$, (c) $r=0.5,\alpha=2,m=0.5$, (d) $r=5,\alpha=1,m=1.5$ (e) $r=5,\alpha=01,m=2$ and (f) $r=5,\alpha=2,m=5$.}
	\end{center}
\end{figure}
\newpage
\noindent \textbf{Theorem 2.} \textit{Let $X  \sim \mathcal{NBRIG}(r,\alpha,m)$ be a negative binomial -reciprocal inverse Gaussian distribution as defined in (\ref{sr}) then its factorial moment of order $k$ is given by }
\begin{equation} \label{fk} 
\mu_{[k]}(X)=  \frac{\Gamma(r+k)}{\Gamma(r)} \sum_{j=0}^{x}\binom{x}{j}(-1)^{j}\sqrt{\frac{\alpha}{\alpha-2(k-j) }} exp\left\{\frac{\alpha}{m^2}\left[m-\frac{m}{\sqrt{\alpha}}\sqrt{\alpha-2(k-j)}\right]\right\}, 
\end{equation}
\noindent \textit{Proof:} If  $X|\lambda\sim \mathcal{NB}(r,p=e^{-\lambda})$ and  $\lambda\sim \mathcal{RIG}(\alpha,m)$, then factorial moment of order $k$ can be find out by using concept of conditional moments as 
 \begin{equation*}
\mu _{[k]}(x)=E_{\lambda}\left[\mu _{[k]}(x|\lambda)\right]
\end{equation*}
Using the factorial moment of order $k$ of $ \mathcal{NB}(r,p)$ , $\mu _{[k]}(x)$ becomes 
\begin{equation*}
\mu _{[k]}(x)=E_{\lambda}\left[\frac{\Gamma(r+k)}{\Gamma(r)}(e^\lambda -1)^k \right]=\frac{\Gamma(r+k)}{\Gamma(r)}E_{\lambda}(e^\lambda -1)^k
\end{equation*}
Through the  binomial expansion of  $(e^\lambda -1)^k=\sum_{j=0}^{k}\binom{k}{j}(-1)^{j}e^{\lambda(k-j)}$, $\mu _{[k]}(x)$ can be written as 
\begin{equation*}
\begin{aligned}
\mu _{[k]}(x)=&\frac{\Gamma(r+k)}{\Gamma(r)}\sum_{j=0}^{k}\binom{k}{j}(-1)^{j}E_{\lambda}(e^{\lambda(k-j)})\\
=&\frac{\Gamma(r+k)}{\Gamma(r)}\sum_{j=0}^{k}\binom{k}{j}(-1)^{j}M_{\lambda}(k-j)
\end{aligned}
\end{equation*}
From the mgf of $\mathcal{RIG}(\alpha,m)$  given in Equation (\ref{mxt}) with $t=k-j$, we get finally factorial moment of order $k$ as:
\begin{equation*}
\frac{\Gamma(r+k)}{\Gamma(r)} \sum_{j=0}^{x}\binom{x}{j}(-1)^{j}\sqrt{\frac{\alpha}{\alpha-2(k-j) }} exp\left\{\frac{\alpha}{m^2}\left[m-\frac{m}{\sqrt{\alpha}}\sqrt{\alpha-2(k-j)}\right]\right\}
\end{equation*}

which proves the theorem. \hfill $\blacksquare$ 
\par
The mean, second order moment and variance can be obtained directly from (\ref{fk}) which are given by 
\begin{eqnarray} 
E\left(X\right)&=& r\left[M_{\lambda}(1)-1\right], \\
E\left(X^2\right)&=& (r+r^2)M_{\lambda}(2)-(r+2r^2)M_{\lambda}(1)+r^2, \\
V\left(X\right)&=& (r+r^2)M_{\lambda}(2)-rM_{\lambda}(1)-r^2M_{\lambda}^{2}(1), 
\end{eqnarray}
where $M_{\lambda}(u)$ is the mgf of $\mathcal{RIG}(\alpha,m)$  defined in (\ref{mxt}). \par 
Overdispersion($\frac{Variance}{Mean}>1$) is an important property in count data. The next theorem establishes that the negative binomial-reciprocal inverse Gaussian distribution is overdispersed as compared to the negative binomial distribution with the same mean. \par 
\noindent \textbf{Theorem 3.} \textit{Let $X$ be a random variable following  $\mathcal{RIG}(\alpha,m)$  whose pdf is given in Equation (3) and $\tilde{X}$ is another random variable following negative binomial distribution i.e.,$ \tilde{X}\sim \mathcal{NB}(r,p=\left[E(e^\lambda)\right]^{-1})$. Suppose consider another random variable $X$ having negative binomial -reciprocal inverse Gaussian distribution which is defined by stochastic representation given in (\ref{sr}). Then we have:  }
\begin{enumerate}
	\item [(i)] E($\tilde{X}$)=E(X) \& Var(X)> Var($\tilde{X}$).
	\item [(ii)] Var(X)>E(X).
\end{enumerate}

\noindent \textit{Proof:} We have $E(e^\lambda)=M_{\lambda}(1)>1$, then $p=\frac{1}{E(e^\lambda)}$ is well defined. Using the definition of conditional expectation, we have
	\begin{eqnarray}
E(X)&=& E_{\lambda}\left(E(X|\lambda)\right)=r\left(M_{\lambda}(1)-1\right)=r\left[E(e^{\lambda})-1\right], \nonumber \\
Var(X)&=& E_{\lambda}\left[V(X|\lambda)\right]+ V_{\lambda}\left[E(X|\lambda)\right] \nonumber \\
&&= (r+r^{2})M_{\lambda}(2) + rM_{\lambda}(1) -r^{2}M_{\lambda}^{2}(1)  \nonumber \\
&&= rM_{\lambda}(2) + r^{2}M_{\lambda}(2)- rM_{\lambda}(1)- r^{2}M_{\lambda}^{2}(1)  \nonumber \\
&&=  rE\left[e^{2\lambda}\right]+ r^{2}E\left[e^{2\lambda}\right]-rE\left[e^{\lambda}\right]- r^{2} \left(E\left[e^{\lambda}\right] \right)^{2}\nonumber \\
&&=  rE\left[e^{2\lambda}\right] + r^{2}V(e^\lambda)-rE\left[e^{\lambda}\right] \nonumber \\
Var(X)&=&r\left[E(e^{2\lambda})-E(e^\lambda)\right] +r^2 V(e^\lambda)
\end{eqnarray}
Also, since $\tilde{X} \sim \mathcal{NB}(r,p=\left[E(e^\lambda)\right]^{-1}) $, we have 
\begin{equation*}
E(\tilde{X})=r\left[E(e^\lambda)-1\right]=E(X)
\end{equation*}
and 
\begin{equation*}
Var(\tilde{X})=r\left[E(e^\lambda)-1\right]E(e^\lambda)
\end{equation*}
Now, using Equation(14), we obtain,
\begin{equation*}
\begin{split}
Var(X)-Var(\tilde{X})=& r\left[E(e^{2\lambda})-E(e^\lambda)\right] +r^2 V(e^\lambda)- Var(\tilde{X}) \\
=& r\left[E(e^{2\lambda})-E(e^\lambda)\right] +r^2 V(e^\lambda)- r\left[E(e^\lambda)-1\right]E(e^\lambda)  \\
=& rE(e^{2\lambda})-rE(e^\lambda) +r^2 V(e^\lambda)- r\left(E(e^\lambda) \right)^2+rE(e^\lambda) \\
=&(r+r^2 )V(e^\lambda)>0 \\
\end{split}
\end{equation*}
It follows  that
\begin{equation}
 Var(X)>Var(\tilde{X})
\end{equation}

(ii) Since $\tilde{X} \sim \mathcal{NB}(r,p=\left[E(e^\lambda)\right]^{-1}) $ \\
$\Rightarrow Var(\tilde{X})>E(\tilde{X})$, but $E(\tilde{X})=E(X)$ 
\begin{equation}
\Rightarrow  Var(\tilde{X})>E(X)
\end{equation}
Combining (15) and (16), it follows that $Var(X)>E(X)$, \\
which proves the theorem. \hfill $\blacksquare$ 
\section{Collective Risk Model under negative binomial-reciprocal inverse Gaussian distribution}

	In non-life Insurance portfolio, the aggregate loss(S) is a random variable defined as the sum of claims occurred in a certain period of time. Let us consider \begin{equation}
	S=X_1 +X_2 +\cdots+X_N,
	\end{equation}
	where $S$ denote  aggregate losses associated with a set of $N$ observed claims, $X_1 , X_2,\cdots,X_N$ satisfying independent assumptions:
	\begin{enumerate}
		\item The $X_{j's} (j=1,2,\cdots,N)$ are independent and identically distributed (i.i.d) random variables with cumulative distribution function $F(x)$ and probability distribution function $f(x)$.
		\item The random variables $N,X_1 , X_2,\cdots$ are mutually independent. 
	\end{enumerate} \par 
Here $N$ be the claim count variable representing number of claims in certian time period and $X_{j} :j=1,2,\cdots$ be the amount of jth claim (or claim severity). When  $ \mathcal{NBRIG}(r,\alpha,m)$ is chosen as primary distribution(N), the distribution of aggregate claim $S$ is called compound negative binomial-reciprocal inverse Gaussian distribution $(\mathcal{CNBRIG})$ whose cdf is given by 
\begin{equation*}
\begin{split}
F_{S}(x)=& P(S\leq x) \\
=& \sum_{n=0}^{\infty}p_{n}P(S\leq x|N=n) \\
=& \sum_{n=0}^{\infty}p_{n}F_{X}^{\star n}(x) \\
\end{split}
\end{equation*}
where $F_{X}(x)=P(X\leq x)$ is the common distribution of $X_{j}^{,}s$ and $p_{n}=P(N=n)$ is given by (\ref{pmf}). $F_{X}^{\star }(n)$ is the n-fold convolution of the cdf of $X$. It can be obtained as 
\begin{equation*}
F_{X}^{\star 0}(x)=\begin{cases}
0;x<0 \\
1; x\geq 0
\end{cases}
\end{equation*}
Next, we will obtain the recursive formula for the probability mass function of   $ \mathcal{NBRIG}(r,\alpha,m)$ distribution in the form of a theorem. 

\noindent \textbf{Theorem 4.} \textit{Let $p(k;r)$ denote the probability mass function (PMF) of an  $\mathcal{NBRIG}(r,\alpha,m)$ and for $r=1,2,\cdots$, the expression for recursive formula is:}
\begin{equation} \label{rcr} 
p(k;r)=\frac{r+k-1}{k}\left[p(k-1;r)-\frac{r}{r+k-1}p(k-1;r+1)\right], 
\end{equation}
\textit{with $ k=1,2,\cdots.$}\\
\noindent \textit{Proof:} The PMF of negative binomial distribution can be written as 
\begin{equation*}
p(k|\lambda)=\binom{r+k-1}{k}e^{-\lambda r}(1-e^{-\lambda})^{k} \quad ; k=0,1,\cdots
\end{equation*}
Now,
\begin{equation*}
\begin{split}
\frac{p(k|\lambda)}{p(k-1|\lambda)}=& \frac{\binom{r+k-1}{k}e^{-\lambda r}(1-e^{-\lambda})^{k}}{\binom{r+k-2}{k-1}e^{-\lambda r}(1-e^{-\lambda})^{k-1}}\\
=& \frac{r+k-1}{k}(1-e^{-\lambda}) \\
\frac{p(z=k|\lambda)}{p(z=k-1|\lambda)}=& \frac{r+k-1}{k}(1-e^{-\lambda}), \quad; k=1,2,\cdots. \\
\end{split}
\end{equation*}
\begin{equation} \label{rucr}
p(z=k|\lambda)=p(z=k-1|\lambda)\frac{r+k-1}{k}(1-e^{-\lambda}) ,\quad k=1,2,\cdots.
\end{equation}
Using the definition of $\mathcal{NBRIG}(r,\alpha,m)$ and (\ref{rucr}),  we have:
\begin{equation*}
\begin{split}
p(k|r)=& \int_{0}^{\infty}p(z=k|\lambda)f(\lambda)d\lambda\\
=&\int_{0}^{\infty} \frac{r+k-1}{k}(1-e^{-\lambda})p(z=k-1|\lambda)f(\lambda)d\lambda \\
= &\int_{0}^{\infty} \frac{r+k-1}{k} p(z=k-1|\lambda)f(\lambda)d\lambda-\int_{0}^{\infty} \frac{r+k-1}{k} p(z=k-1|\lambda)e^{-\lambda}f(\lambda)d\lambda\\
=& \frac{r+k-1}{k}p(k-1;r)-\frac{r+k-1}{k}\int_{0}^{\infty}e^{-\lambda}p(z=k-1|\lambda)f(\lambda)d\lambda .\\
\end{split}
\end{equation*}
Also, we obtain now
\begin{equation*}
\begin{split}
& \int_{0}^{\infty}e^{-\lambda}p(z=k-1|\lambda)f(\lambda)d\lambda\\
=&\int_{0}^{\infty} e^{-\lambda}\binom{r+k-2}{k-1}e^{-\lambda r}(1-e^{-\lambda})^{k-1}f(\lambda)d\lambda \\
= & \frac{r}{r+k-1}\int_{0}^{\infty} \binom{r+1+k-2}{k-1}e^{-\lambda(r+1)}(1-e^{-\lambda})^{k-1} f(\lambda)d\lambda\\
=& \frac{r}{r+k-1}p(k-1;r+1) , \\
\end{split}
\end{equation*}
and thus (\ref{rcr}) is obtained. \hfill $\blacksquare$ \\
\noindent \textbf{Theorem 5.} \textit{If the claim sizes are absolutely continuous random variables with pdf $f(x)$ for $x>0$, then the pdf $g_{s}(x;r)$  of the $(\mathcal{CNBRIG})$ satisfies the integral equation:}
	\begin{eqnarray} \label{ige} 
g_{s}(x;r)&=& p(0;r)+\int_{0}^{x}\frac{ry+x-y}{x}g_{s}(x-y;r)f(y)dy \nonumber \\
&& -  \int_{0}^{x}\frac{ry}{x}g_{s}(x-y;r+1)f(y)dy.
\end{eqnarray}
\noindent \textit{Proof:} The aggregate claim distribution is given by  
	\begin{eqnarray} \label{ige} 
g_{s}(x;r)&=& \sum_{k=0}^{\infty}p(k;r)f^{k \star}(x) \nonumber \\
&& =  p(0;r) f^{0 \star}(x)+ \sum_{k=1}^{\infty}p(k;r)f^{k \star}(x)\nonumber 
\end{eqnarray}
Using (\ref{rcr}), we get:
	\begin{eqnarray} \label{gsr}
g_{s}(x;r)&=& p(0;r) + \sum_{k=1}^{\infty}f^{k \star}(x) \left[\frac{r+k-1}{k}\left(p(k-1;r)-\frac{r}{r+k-1}p(k-1;r+1)\right)\right]\nonumber  \\
&& =  p(0;r) + \sum_{k=1}^{\infty}\frac{r-1}{k}p(k-1;r)f^{k \star}(x)+\sum_{k=1}^{\infty}p(k-1;r)f^{k \star}(x) \nonumber \\
&& + \sum_{k=1}^{\infty}\frac{r}{k}p(k-1;r+1)f^{k \star}(x)  \nonumber \\
\end{eqnarray} 
Using the identities:
\begin{eqnarray} 
f^{k \star}(x)&=& \int_{0}^{x} f^{(k-1) \star}(x-y)f(y)dy, \quad  k=1,2,\cdots  \\
\frac{f^{k \star}(x)}{k}&=& \int_{0}^{x}\frac{y}{x} f^{(k-1) \star}(x-y)f(y)dy, \quad k=1,2,\cdots 
\end{eqnarray}
Therefore, now (\ref{gsr}) can be written as:
\begin{eqnarray}  \label{gssr}
&&  \sum_{k=1}^{\infty}(r-1)p(k-1;r)\int_{0}^{x}\frac{y}{x} f^{(k-1) \star}(x-y)f(y)dy \nonumber \\
&&+ \sum_{k=1}^{\infty}p(k-1;r)\int_{0}^{x}f^{(k-1) \star}(x-y)f(y)dy\nonumber  \\
&&- \sum_{k=1}^{\infty}rp(k-1;r+1)\int_{0}^{x}\frac{y}{x} f^{(k-1) \star}(x-y)f(y)dy\nonumber  \\
&=& \int_{0}^{x}\frac{ry+x-y}{x} f^{(k-1) \star}(x-y)f(y)dy \sum_{k=1}^{\infty}p(k-1;r) \nonumber \\
&&-  \int_{0}^{x}\frac{ry}{x} f^{(k-1) \star}(x-y)f(y)dy \sum_{k=1}^{\infty}p(k-1;r+1) 
\end{eqnarray} 
Also we can write:
\begin{equation*}
\begin{split}
& g_{s}(x,r)=\sum_{k=1}^{\infty}p(k-1;r)f^{(k-1) \star}(x) , \quad k=1,2,\cdots\\
& g_{s}(x-y,r)=\sum_{k=1}^{\infty}p(k-1;r)f^{(k-1) \star}(x-y) \\
& g_{s}(x-y,r+1)=\sum_{k=1}^{\infty}p(k-1;r+1)f^{(k-1) \star}(x-y)  \\
\end{split}
\end{equation*}
Thus (\ref{gssr}) becomes:
\begin{equation*}
\begin{split}
& \int_{0}^{x}\frac{ry+x-y}{x}f(y)dy g_{s}(x-y,r)-\int_{0}^{x}\frac{ry}{x}f(y)dy g_{s}(x-y,r+1) \\
\end{split}
\end{equation*}
Therefore we finally get:
	\begin{eqnarray*} 
g_{s}(x;r)&=& p(0;r)+\int_{0}^{x}\frac{ry+x-y}{x}g_{s}(x-y;r)f(y)dy \nonumber \\
&& -  \int_{0}^{x}\frac{ry}{x}g_{s}(x-y;r+1)f(y)dy,
\end{eqnarray*}\\
Hence proved.  \hfill $\blacksquare$ \\
The Integral equation obtained in above theorem can be solved numerically  in practice and the discrete version of it can be obtained in a similar fashion by interchanging $\int_{0}^{x}$ to $ \sum_{y=1}^{x}$ in expressions (22) and (23) (Rolski et al. (1999)). So its discrete version obtained are as 
	\begin{eqnarray*} 
	g_{s}(x;r)&=& p(0;r)+\sum_{y=1}^{x}\frac{ry+x-y}{x}g_{s}(x-y;r)f(y) \nonumber \\
	&& -  \sum_{y=1}^{x}\frac{ry}{x}g_{s}(x-y;r+1)f(y).
\end{eqnarray*}
\section{Multivariate version of negative binomial-reciprocal inverse Gaussian distribution}
In this section, we propose the multivariate version of negative binomial-reciprocal inverse Gaussian distribution which is actually extension of definition (\ref{sr}). The multivariate negative binomial- reciprocal inverse Gaussian distribution can be considered as a mixture of independent  $\mathcal{NB}(r_i ,p=e^{-\lambda}), i=1,2,\cdots,d$ combined with a reciprocal Gaussian distribution.\\
\noindent \textbf{Definition 2.} A multivariate negative binomial-reciprocal inverse Gaussian  distribution $(X_{1}, X_{2},\cdots,X_{d})$ is defined by  stochastic representation:
\begin{equation*}  
\begin{aligned}
X_{i}|\lambda\sim  &\mathcal{NB}(r_{i},e^{-\lambda}),\quad i=1,2,\cdots,d\; are \; independent\\
\& \quad\lambda \sim  &\mathcal{RIG} (\alpha,m)
\end{aligned} 
\end{equation*}
Using the same arguments as mentioned in section 2, the joint PMF obtained is given by:
\begin{equation}   \label{mv}
\begin{aligned}
P(X_1=x_1 ,X_2 =x_2 ,\cdots,X_d =x_d)= & \prod_{i=1}^{d}\binom{r_i  +x_{i}-1}{x_i}\sum_{j=0}^{\tilde{x}}(-1)^j \binom{\tilde{x}}{j}\\
& \times \sqrt{\frac{\alpha}{\alpha+2(\tilde{r}+j) }} exp\left\{\frac{\alpha}{m^2}\left[m-\frac{m}{\sqrt{\alpha}}\sqrt{\alpha+2(\tilde{r}+j)}\right]\right\}
\end{aligned} 
\end{equation}
where $x_{1},x_{2},\cdots,x_{d}=0, 1, 2,\cdots; \alpha, m, r_1 , r_2 , \cdots, r_d >0$ and 
\begin{eqnarray} 
\tilde{r}=r_{1}+ r_{2}+\cdots+r_{d}, \\
\tilde{x}=x_{1}+ x_{2}+\cdots+x_{d}. 
\end{eqnarray}
The above joint PMF can be written in a more convenient form for the purpose of computing  multivariate probabilities. Let $\tilde{Y} \sim \mathcal{NBRIG}(\tilde{r},\alpha,m)$, where  $\tilde{r}$ is given in (26), an alternative structure for (\ref{mv}) with $d\geq 2$ is given by: 
\begin{eqnarray}  \label{mvv}
&&P(X_1=x_1 ,X_2 =x_2 ,\cdots,X_d =x_d) \nonumber \\  
&=& \frac{\prod_{i=1}^{d}\binom{r_i  +x_{i}-1}{x_i}}{\binom{\tilde{r}  +\tilde{x}-1}{\tilde{x}}}\cdot P(\tilde{Y}=\tilde{x})
\end{eqnarray} 
where $\tilde{x}$ is defined in equation (27). The marginal distribution will be obviously as $\tilde{X} \sim \mathcal{NBRIG}(r_i ,\alpha,m)$, $i=1,2,\cdots, d$ and any subvector $(X_{1},X_{2},\cdots,X_{s})$ with $s<d$ is again a multivariate negative binomial- reciprocal inverse Gaussian distribution of dimension $s$. Using (11) and (13), the following expressions for moments can be obtained as:
\begin{eqnarray} 
E\left(X\right)&=& r_{i}\left[M_{\lambda}(1)-1\right],\quad i=1,2,\cdots,r \\
V\left(X\right)&=& (r_{i}+r_{i}^2)M_{\lambda}(2)-r_{i}M_{\lambda}(1)-r_{i}^2M_{\lambda}^{2}(1), i=1,2,\cdots,r \\
Cov\left(X_{i},X_{j}\right) &=& r_{i}r_{j}\left[M_{\lambda}(2)-M_{\lambda}^{2}(1)\right], \quad i\neq j
\end{eqnarray}
Since $M_{\lambda}(2)=E\left[e^{2\lambda}\right]$ \&  $M_{\lambda}(1)=E\left[e^{\lambda}\right]$ $\Rightarrow V(e^\lambda)=M_{\lambda}(2)-M_{\lambda}^{2}(1)$. \\ Therefore $Cov\left(X_{i},X_{j}\right)=r_{i}r_{j}V(e^\lambda); i\neq j$ \\
Now $\rho(X_{i},X_{j})=\frac{Cov\left(X_{i},X_{j}\right)}{\sigma_{x_{i},x_{j}}}=\frac{r_{i}r_{j}V(e^\lambda)}{\sigma_{x_{i},x_{j}}}>0$, \\
Thus, it follows $\rho(X_{i},X_{j})>0$.
\section{Estimation}
	In this Section, we will discuss one of the popular method of estimation namely Maximum Likelihood Estimation (MLE) for the estimation of the parameters of $\mathbb{NBRIG}(r,\alpha,m)$ distribution. Suppose $\underline{\mathbf{x}}=\lbrace x_{1}, x_{2},\cdots,x_{n} \rbrace$  be a random sample of size $n$ from the  $\mathbb{NBRIG}(r,\alpha,m)$ distribution with PMF given in (\ref{pmf}). The likelihood function is given by 
\begin{equation} \label{likeli} 
L(m,\alpha,r|\underline{\mathbf{x}})=\prod\limits_{i=1}^{n}\binom{r+ x-1}{x}\sum_{j=0}^{x}\binom{x}{j}(-1)^{j}\sqrt{\frac{\alpha}{\alpha+2(r+j) }} exp\left\{\frac{\alpha}{m^2}\left[m-\frac{m}{\sqrt{\alpha}}\sqrt{\alpha+2(r+j)}\right]\right\}
\end{equation}
	The log-likelihood function corresponding  to (\ref{likeli}) is obtained as 
\begin{equation}
\begin{split}
\log L(m,\alpha,r|\underline{\mathbf{x}})=&\sum_{i=1}^{n} \log\binom{r+ x_{i}-1}{x_{i}}+ \sum_{i=1}^{n}\log\left[\sum_{j=1}^{x_{i}}\binom{x_{i}}{j}(-1)^j \sqrt{\frac{\alpha}{\alpha+2(r+j)}}\right] \\
 &+\sum_{i=1}^{n}\frac{\alpha}{m^2}\left[m-m\sqrt{\frac{\alpha+2(r+j)}{\alpha}}\right] 
\end{split}
\end{equation}
	The ML Estimates  $\hat{m}$ of $m$, $\hat{\alpha}$ of $\alpha$ and $\hat{r}$ of $r$, respectively, can be obtained by solving equations 
\begin{equation*}
 \frac{\partial \log L}{\partial m}=0, \quad \frac{\partial \log L}{\partial \alpha }=0 \quad  \text{and} \quad  	\frac{\partial \log L}{\partial r}=0.
\end{equation*}
where 
\begin{equation}
\begin{split}
\frac{\partial \log L}{\partial m}=&\frac{n \alpha   \left(\sqrt{\frac{\alpha +2 j+2 r}{\alpha }}-1\right)}{m^2} ,
\end{split}
\end{equation}
\begin{equation}
\begin{split}
\frac{\partial \log L}{\partial \alpha}=&\sum _{i=1}^n \frac{\sum _{j=0}^{x_i} \frac{(-1)^j \binom{x_i}{j} \left(\frac{1}{\alpha +2 (j+r)}-\frac{\alpha }{(\alpha +2 (j+r))^2}\right)}{2 \sqrt{\frac{\alpha }{\alpha +2 (j+r)}}}}{\sum _{j=0}^{x_i} (-1)^j \binom{x_i}{j} \sqrt{\frac{\alpha }{\alpha +2 (j+r)}}} ,
\end{split}
\end{equation}
\begin{equation}
\begin{split}
\frac{\partial \log L}{\partial r}=&\sum _{i=1}^n \frac{\sum _{j=0}^{x_i} -\frac{\alpha  (-1)^j \binom{x_i}{j}}{\sqrt{\frac{\alpha }{\alpha +2 (j+r)}} (\alpha +2 (j+r))^2}}{\sum _{j=0}^{x_i} (-1)^j \binom{x_i}{j} \sqrt{\frac{\alpha }{\alpha +2 (j+r)}}}+n \left(\psi ^{(0)}(r+x)-\psi ^{(0)}(r)\right) ,
\end{split}
\end{equation}
where $\psi(r)=\frac{d}{dr}\Gamma(r)$ is digamma function. As the above  equations  are not in closed form and hence cannot be solved explicitly. So we make use of a suitable  iterative technique to find the ML estimates  numerically.
\section{Numerical Illustrations}
In order to illustrate how the proposed distribution fits the count data, two well-known data sets have been taken into consideration from actuarial literature.

\textbf{Illustration 1:} The first data set contains details about the number of automobile liability policies in Switzerland for private cars: (Klugman et al. (1998), pp.245 and Denuit(1997), pp. 240). For comparison we fit Poisson $(\mathcal{P})$, negative binomial ($\mathcal{NB}$) and negative binomial- reciprocal inverse Gaussian  ($\mathcal{NBRIG}$) distributions to this data set by using technique of maximum likelihood and  estimated parameters involved in each model are given in Table 1. In order to test the goodness of fit, chi-square test criterion has been employed. It is to pertinent to mention that expected frequencies have been grouped into classes for having cell frequencies greater than five in order to apply the chi-square goodness of fit. Based on the results like log likelihood , $p$-value, that there exists enough statistical evidence that the proposed distribution (i.e.$\mathcal{NBRIG}$ ) fits the data very well. Akaike’s information criterion (AIC) have also been used as a measure of model validation. This AIC value of the model is defined by $AIC=2k-2\ln(\hat{l})$ where $k$ is the number of parameters involved in the model and $\hat{l}$ is the maximum value of the likelihood function for the model.  A model with a lower AIC is preferable. Based on this value, the $\mathcal{NBRIG}$ distribution provides the best fit to data (Table 1). \\
\begin{table}[htbp]
	\centering
	\caption{Number of automobile liability policies in Switzerland for private cars}
	\begin{tabular}{rrrrrrr} \hline
		& {Observed} & \multicolumn{5}{c}{Expected frequency} \\ \hline
		Count & Frequency & \multicolumn{1}{c} {$\mathcal{P}$} & \multicolumn{1}{c}{$\mathcal{NB}$}  & \multicolumn{1}{c}{$\mathcal{NBRIG}$}       \\ \hline
		0     & 103704   & 102630   & 103724   & 103710 \\
		1     & 14075   & 15921.9   & 13989.9   & 14054.8 \\
		2     &1766    & 1235.07  & 1857.07   & 1787.35\\
		3     &255   & 63.8694  & 245.194   & 251.933\\
		4     & 45   & 2.47717  & 32.2863   & 40.2211 \\
		5     & 6     & 0.0768617  & 4.24447   & 7.21741 \\
		6     & 2     & 0.00198739  & 0.557389   &1.43701     \\
		> 6   & 0     & 0   & 0   & 0   \\
		\hline 
		Total & 119853   & 119853        & 119853       & 119853               \\ \hline
		& Estimated &  &  &    \\
		& Parameters &  $ \hat{\lambda}=0.15514$ & $\hat{r}=1.03267$  & $\hat{m}=35.8961,$ \\
		& &  & $\hat{p}=0.87$ & $\hat{\alpha}= 61.4973$\\
		& & & & $\hat{r}=3.4$ \\  \hline
		&       &       &       &       &       &         \\
		& log likelihood & -55108.5 & -54615.3 & -54609 \\ 
		& $\chi^2$ (d.f) & 4220.76(2) & 14.85(2) & 1.32(2)   \\ 
		&   p-value & 0 & 0.0005973  &0.5180237  \\ 
		& AIC &110219  & 109234.6 &109224 \\ \hline 
		
	\end{tabular}%
	\label{tab:addlabel}%
\end{table}
\newpage
\textbf{Illustration 2:} The second data set is included in Klugman et al.(2008) and it consists  of the data set of 9,461 automobile insurance policies where the number of accidents of each policy has been recorded. Again, the Poisson  $(\mathcal{P})$, negative binomial ($\mathcal{NB}$) and  negative binomial- reciprocal inverse Gaussian  ($\mathcal{NBRIG}$) distributions  have been fitted to data by maximum likelihood estimation. Observed and expected values together with parameter estimates including log likelihood, chi-square value, $p$-value and AIC are exhibited in  Table 2. Based on the results from  Table 2, it clearly suggests that  our proposed distribution outperforms other two competing models.

\begin{table}[htbp]
	\centering
	\caption{Observed and expected frequencies for the accident data }
	\begin{tabular}{rrrrrrr} \hline
		& {Observed} & \multicolumn{5}{c}{Expected frequency} \\ \hline
		Count & Frequency & \multicolumn{1}{c} {$\mathcal{P}$} & \multicolumn{1}{c}{$\mathcal{NB}$}  & \multicolumn{1}{c}{$\mathcal{NBRIG}$}       \\ \hline
		0     & 7840   & 25528.6   & 27165.8   & 103710 \\
    	1     & 1317    & 8107.98   & 5664.06   & 14054.8 \\
    	2     & 239    & 1287.56  & 1563.35   & 1787.35\\
	    3     & 42    & 136.312  & 466.683   & 251.933\\
	    4     & 14    & 10.8233  & 144.563   & 40.2211 \\
     	5     & 4     & 0.687502  & 45.7569   & 7.21741 \\
		6     & 4    & 0.04       & 14.6889   &   1.43701     \\
		7     & 1    &0.002  & 4.76263   & 0   \\
		8 +    & 0    & 0   & 1.55569   & 0   \\ 
		\hline 
			Total & 9461   & 9461        & 9461    & 9461              \\ \hline
		& Estimated &  &  &    \\
		& Parameters & $ \hat{\lambda}=0.21$ & $\hat{r}=0.70$  & $\hat{m}=17.42$ \\
		& & & $\hat{p}=0.77$ & $\hat{\alpha}= 24.87$ \\
     	& & & & $\hat{r}=2.03$ \\  \hline
		&       &       &       &       &       &         \\
		& log likelihood & -5490.78 & -5348.04 & -5343.05 \\ 
		& $\chi^2$ (d.f) & 16517.71(3) & 32.20(5) & 8.73(2)    \\ 
		&   p-value & 0 & <0.00054  & 0.0171466  \\ 
		& AIC &10983.56  & 10700.08 &10692.1 \\ \hline 
		
	\end{tabular}%
	\label{tab:addlabel}%
\end{table}
\newpage
\section{Conclusion}
In this paper, we introduce a new three-parameter negative binomial-reciprocal inverse Gaussian  distribution, $\mathcal{NBRIG}(r,\alpha,m)$  including its multivariate extension as well.  This distribution is obtained by mixing the $\mathcal{NB}$ with the $\mathcal{RIG}(\alpha,m)$ distribution. In addition, the moments of the $\mathcal{NBRIG}(r,\alpha,m)$ distribution which includes the factorial moments, mean, variance, are derived. The over-dispersion property has also  been established by a theorem. Moreover, the parameters have been estimated by using the maximum likelihood estimation method. We include an application of the $\mathcal{NBRIG}(r,\alpha,m)$ distribution to fit two real data sets. We found that these two data sets gives best fit with $\mathcal{NBRIG}(r,\alpha,m)$ distribution among the three distributions. We are  hopeful that $\mathcal{NBRIG}(r,\alpha,m)$ distribution may attract wider applications in analyzing count data.

\end{document}